\documentstyle[amssymb,preprint,aps]{revtex}

\begin{document}
\title{The Cangemi-Jackiw manifold in high dimensions and symplectic structure}
\author{L. M. Abreu\thanks{%
E-mail: lma@ufba.br}, A. E. Santana\thanks{%
E-mail: santana@fis.ufba.br}, A. Ribeiro Filho\thanks{%
E-mail: ribfilho@ufba.br}}
\address{${}$Instituto de F\'\i sica, Universidade Federal da Bahia, \\
Campus de Ondina, 40210-340, Salvador, Bahia, Brazil}
\maketitle

\begin{abstract}
The notion of Poincar\'{e} gauge manifold (${\cal G}$), proposed in the
context of an $(1+1)$ gravitational theory by Cangemi and Jackiw ( D.
Cangemi and R. Jackiw, Ann. Phys. (N.Y.) {\bf 225} (1993) 229), is explored
from a geometrical point of view. First ${\cal G}$ is defined for arbitrary
dimensions and in the sequence a symplectic structure is attached to $%
T^{\ast }{\cal G}$. Treating then the case of 5-dimensions, a $(4,1)$
de-Sitter space, applications are presented studing representations of the
Poincar\'{e} group in association with kinetic theory and the Weyl operators
in phase space. The central extension in the Aghassi-Roman-Santilli group
(Aghassi, Roman and Santilli, Phys. Rev. D {\bf 1 }(1970) 2753) is derived
as a subgroup of linear transformations in ${\cal G}$ with 6 dimensions.
\end{abstract}

\section{Introduction}

This paper is dedicated to an investigation of the manifold (to be denoted
by ${\cal G}$ ) proposed by Cangemi and Jackiw (CJ) \cite{cj1,cj2,cj3} in
the study of an ($1+1$)-dimensional gravity as a gauge theory. In the CJ
formalism, the gauge group is given by the generators of the Lie algebra $%
[x_{1},x_{2}]=x_{4},\,\,\,[x_{3},x_{2}]=x_{1},\,\,\,[x_{3},x_{1}]=x_{2},$
which is a central extension, specified by $x_{4}$, of the ($1+1$)-Poincar%
\'{e} group. This Lie algebra is in correspondence to a symmetric bilinear
invariant form given by 
\begin{equation}
\eta =\left( 
\begin{array}{ccc}
g^{\mu \nu } & 0 & 0 \\ 
0 & 0 & -1 \\ 
0 & -1 & 0
\end{array}
\right) ,  \label{m1}
\end{equation}
where $g^{\mu \nu }$ is the $(1+1)$-Minkowski metric, with $diag(g^{\mu \nu
})=(-1,1).$ If, in particular, $g^{\mu \nu }$ is the Euclidean metric, in
which $diag(g^{\mu \nu })=(1,1),$ then \ $\eta $ is the light-cone metric 
\cite{Bazeia,Susskind}.

Among different applications of this approach \cite{wzw,olive,sf,luk}, an
analysis of $\eta $-like metrics and associated Lie algebras \cite{esd1} has
shown that this correspondence is not unique, and that in five dimensions $%
\eta ,$ with $g^{\mu \nu }$ being the Euclidean metric in three dimensions,
can be used to write the relativistic and the non-relativistic physics in a
geometric unified way. One interesting result in this analysis is that the
Galilean physics can be derived as a manifestly covariant theory \cite
{kun1,kun2,kun3,kun4,tak1,tak2,tak3,tak4}. Such results point to the
richness of ${\cal G}$ and for the need of other additional studies, as for
instance by exploring higher dimensions and the symplectic structure
attached to $T^{\ast }{\cal G}$. These aspects are addressed here.

To begin with, in Section 2, the CJ manifold is defined in $N$ dimensions
and then the linear group of transformations in this $N$ dimensional space
is studied. The notion of symplectic 2-form is introduced in Section 3.
Therefore (canonical) representations of Lie algebras in phase space,
including representations with generalized van Hove operators \cite{van Hove}%
, are analyzed. In Section 4, considering ${\cal G}$ in 5-dimensions and
using the notion of embedding (following along the lines established in Ref. 
\cite{esd1}), a relativistic kinetic equation is derived and studied. This
equation, which is but an example of the classical thermofield dynamics
formalism \cite{marco1}, has a local collision term obtained from symmetry
elements via a Lagrangian\ written in phase space. A particular solution
leads then to a wave-travelling distribution function. The symplectic
structure is also used with the Weyl product (first introduced in the
context of the Wigner-function picture of quantum mechanics \cite{Wigner}),
and applied to the analysis of a scheme proposed by McDonald and Kaufman 
\cite{mck} to derive kinetic equations governing the behavior of the
electromagnetic waves in a dispersive medium. The main result here is\ to
show that the symplectic structure associated to $T^{\ast }{\cal G}$\
provides a geometric interpretation for the McDonald and Kaufman method.

In Section 5 ${\cal G}$ is studied in 6-dimensions. In this case we show
that a subgroup of the linear transformations in ${\cal G}$ is a general set
of affine transformations in the $(3+1)$-Minkowski space with a central
charge. This kind of group was studied by Aghassi, Roman and Santilli $\ $ 
\cite{Santilli1,Santilli3,Santilli2}, motived by the\ following problem. In
the contraction of the Poincar\'{e} group, taking $c\rightarrow \infty ,$
the extended Galilei group is recovered but with mass interpreted as a
central charge and not as the value of a Casimir invariant, as it is in the
case of the Poincar\'{e} Lie algebra. Analyzing this kind of difficulty
(representing a problem at least at the level of rigor), Aghassi, Roman and
Santilli considered a general set of linear transformations in the $(3+1)$%
-Minkowski space including, \ beyond the Lorentz rotations and the
translations, a boost given by $b^{\mu }u,$ where $b^{\mu }$\ is a
4-velocity and $u$ is a proper-time scalar. In this scenario, mass arises
via a central extension of this group, which is a kind of Galilean
relativistic symmetry. In our analysis, using the CJ manifold, without the
use of central extension, mass emerges as a Casimir invariant in the Lie
algebra of the linear group of transformations in the 6-dimensional ${\cal G}
$ space. The connection with the usual field theory is discussed via the
scalar representation. Final concluding remarks are presented in Section 6.

\section{The CJ manifold and tensor fields}

\bigskip Let $\left( {\cal G}{\sl ,\eta }\right) $ be a $N$-dimensional
pseudo-Riemannian manifold, where the metric $\eta $ is given by 
\begin{equation}
\eta =\eta _{\mu \nu }e^{\mu }\otimes e^{\nu },  \label{met}
\end{equation}
with $\left\{ e^{\mu }\right\} $ being basis of cotangent space of ${\cal G}$
at arbitrary point $p\in {\cal G}$; the components of $\eta ,$ say $\eta
_{\mu \nu },$ are represented by the matrix 
\begin{equation}
\left( \eta _{\mu \nu }\right) =\left( 
\begin{array}{ccccc}
{\bf I}_{ij} & \hspace{0.06in}0\hspace{0.05in} & \hspace{0.03in}0 & \ldots & 
\hspace{0.03in}0 \\ 
0 & {\Bbb B}_{1} & 0 & \ldots & 0 \\ 
0 & 0 & {\Bbb B}_{2} & \ldots & 0 \\ 
\vdots & \vdots & \vdots & \ddots & \vdots \\ 
0 & 0 & 0 & \ldots & {\Bbb B}_{\frac{N-k}{2}}
\end{array}
\right) _{N,N},  \label{mat}
\end{equation}
where ${\bf I}_{ij}$ is the $k$-dimensional Euclidean metric, $diag\left( 
{\bf I}_{ij}\right) =\left( 1,1,...,1\right) $ (with $i,j\leq k$)$,$ and 
\begin{equation}
{\Bbb B}_{i}=\left( 
\begin{array}{cc}
0 & -1 \\ 
-1 & 0
\end{array}
\right) .  \label{block}
\end{equation}
{\bf \ }The restriction is that $N-k$ is a even number.

Consider two arbitrary vectors $X$ and $Y$ in ${\cal G}$, which are written
in terms of the basis $\left\{ e_{\mu }\right\} $ of $T_{p}{\cal G}$ in the
form $X=X^{\mu }e_{\mu }$ and $Y=Y^{\mu }e_{\mu }$. Using Eq.(\ref{mat}),
the inner product is 
\begin{eqnarray}
\left( X\mid Y\right) &=&\eta \left( X,Y\right)  \nonumber \\
&=&\sum_{i=1}^{k}X^{i}Y^{i}-\sum_{i=1}^{\frac{N-k}{2}}\left(
X^{k+2i}Y^{k+2i-1}+X^{k+2i-1}Y^{k+2i}\right) .  \label{inner}
\end{eqnarray}
The rules raising and lowering indices, determined by $X^{\mu }=\eta ^{\mu
\nu }X_{\nu }$ and $X_{\mu }=\eta _{\mu \nu }X^{\nu }$with $\eta _{\mu \nu
}=\eta ^{\mu \nu },$ give rise to the following relations, 
\begin{eqnarray}
X^{i} &=&X_{i},\text{ if }1\leq i\leq k;  \label{ab1} \\
X^{\mu } &=&X^{k+j}=-X_{\mu +a},\text{ if \ }k+1\leq \mu \leq N;a=\pm 1,
\label{ab2}
\end{eqnarray}
where the parameter $a$ is $+1$ if $j$ is odd or $-1$ if $j$ is even.

Since that in Eq. (\ref{met}) the idea of a $(q=0,r=2)$-tensor at $p\in 
{\cal G}$ is introduced, we can now generalize it for an arbitrary tensor at 
$p$, say ${\cal J}_{r,p}^{q}\left( {\cal G}\right) ,$ by the form 
\[
T:T_{p}^{\ast }{\cal G}^{\left( 1\right) }\otimes ...\otimes T_{p}^{\ast }%
{\cal G}^{\left( q\right) }\otimes T_{p}{\cal G}^{\left( 1\right) }\otimes
...\otimes T_{p}{\cal G}^{\left( r\right) }\longmapsto {\cal R}, 
\]
such that $T$ is written in terms of the bases as 
\begin{equation}
T=T_{\;\;\;\;\;\;\;\nu _{1}...\nu _{r}}^{\mu _{1}...\mu _{q}}e_{\mu
_{1}}\otimes ...\otimes e_{\mu _{q}}\otimes e^{\nu _{1}}\otimes ...\otimes
e^{\nu _{r}},  \label{tensor}
\end{equation}
where $T_{\;\;\;\;\;\;\;\nu _{1}...\nu _{r}}^{\mu _{1}...\mu _{q}}=T\left(
e^{\mu _{1}},...,e^{\mu _{q}},e_{\nu _{1}},...,e_{\nu _{r}}\right) $.

It is interesting at this point \ to analyze linear transformations in $%
{\cal G}$ of type 
\begin{equation}
\overline{x}^{\mu }=G_{\;\nu }^{\mu }x^{\nu }+a^{\mu },  \label{transf}
\end{equation}
with $\left| G\right| =1$. Considering infinitesimal transformations, such
that $G_{\;\nu }^{\mu }=\delta _{\;\nu }^{\mu }+\epsilon _{\;\nu }^{\mu },$
we can see that the generators of these transformations are given by the
operators 
\begin{eqnarray}
M_{\mu \nu } &:&=-i\left( x_{\mu }\partial _{\nu }-x_{\nu }\partial _{\mu
}\right) ,  \label{op mom ang} \\
P_{\mu } &:&=-i\partial _{\mu },  \label{op mom}
\end{eqnarray}
which are defined in the space of functions at $p\in {\cal G}$. These
generators satisfy the Lie algebra,

\begin{eqnarray}
\left[ M_{\mu \nu },M_{\rho \sigma }\right] &=&-i\left( \eta _{\nu \rho
}M_{\mu \sigma }-\eta _{\mu \rho }M_{\nu \sigma }+\eta _{\mu \sigma }M_{\nu
\rho }-\eta _{\nu \sigma }M_{\mu \rho }\right) ,  \label{comm1} \\
\left[ P_{\mu },M_{\rho \sigma }\right] &=&-i\left( \eta _{\mu \rho
}P_{\sigma }-\eta _{\mu \sigma }P_{\rho }\right) ,  \label{comm2} \\
\left[ P_{\mu },P_{\nu }\right] &=&0.  \label{comm3}
\end{eqnarray}
Eqs. (\ref{comm1})-(\ref{comm3}) will be called a $g$-Lie algebra.

For the case of 5-dimensions, in order to rederive the results of Ref.\cite
{esd1}, we introduce the notation, 
\begin{equation}
J_{lm}=M_{lm}=-i\left( x^{l}\partial _{m}-x^{m}\partial _{l}\right) ,
\label{mom ang}
\end{equation}
being $1\leq l,m\leq k$. We also have 
\begin{equation}
G_{lm}=M_{k+l,m}=i\left( x^{k+l+a}\partial _{m}+x^{m}\partial _{k+l}\right) ,
\label{boost}
\end{equation}
if $1\leq l\leq N-k$ and $1\leq m\leq k$ ($a=\pm 1$ if $l$ is odd or even,
respectively), and finally 
\begin{equation}
D_{lm}=M_{k+l,k+m}=i\left( x^{k+l+a_{l}}\partial
_{k+m}-x^{k+m+a_{m}}\partial _{k+l}\right) ,  \label{temp}
\end{equation}
if $1\leq l,m\leq N-k$ ($a_{l},a_{m}=\pm 1$ if $l,m$ are odd or even,
respectively).

For the sake of applications in next sections, it is interesting to observe
the nature of different embeddings from the Euclidean space ${\cal E}$ into $%
{\cal G}$. Let us then exemplify two particular embeddings.

The first case of embedding is defined by 
\begin{equation}
\Im _{1}:{\bf A}\longmapsto A=\left( {\bf A,}\frac{A^{k+1}}{\sqrt{2}},\frac{%
A^{k+1}}{\sqrt{2}},\ldots ,\frac{A^{k+\frac{N-k}{2}}}{\sqrt{2}},\frac{A^{k+%
\frac{N+k}{2}}}{\sqrt{2}}\right) ,  \label{emb1}
\end{equation}
where ${\bf A}$ is a vector in the $k$-dimensional Euclidean space(${\cal E}%
^{k}$). Therefore, from the Eq. (\ref{inner}), the norm of this class of
vectors in ${\cal G}$ is 
\begin{equation}
\left\| A\right\| ^{2}\stackrel{\Im _{1}}{=}{\bf A}^{2}-\sum_{i=1}^{\frac{N-k%
}{2}}\left( A^{k+i}\right) ^{2}.  \label{norm2}
\end{equation}
We can have vectors of type $\left\| A\right\| ^{2}=0$ or $\left\| A\right\|
^{2}\gtrless 0$. Hence this embedding gives rise to a Minkowski space in ($%
N-1)$-dimensions, which is a hyperplane in ${\cal G}$. In other words 
\begin{equation}
A^{\mu }A_{\mu }=\eta _{\mu \nu }A^{\mu }A^{\nu }\stackrel{\Im _{1}}{%
\longmapsto }A^{\mu }A_{\mu }=g_{\mu \nu }A^{\mu }A^{\nu },  \label{mink11}
\end{equation}
where $g_{\mu \nu }$ is just the Minkowski metric, given by $diag\left(
g_{\mu \nu }\right) =\left( -1,1,\ldots ,1\right) $ which has $N-1$ entries
and $A^{k+1}\equiv $ $A^{0}$. Then the $5$-dim vectors become $A=\left(
A^{0},{\bf A}\right) .$ We will use the embedding $\Im _{1}$ to study
representations of the Poincar\'{e} Group in phase space.

The second case of embedding to be considered is the following 
\begin{equation}
\Im _{2}:{\bf A}\longmapsto A=\left( {\bf A,}A^{k+1},0,A^{k+2},0,\ldots
,A^{k+\frac{N-k}{2}},0\right) .  \label{emb2}
\end{equation}
which results in vectors with norm $(A|A)={\eta }_{\mu \nu }A^{\mu }A^{\nu }%
\stackrel{\Im _{2}}{=}{\bf A}^{2}.$

\section{Symplectic structures in $T^{\ast }{\cal G}$}

Let us define a $2N$-dimensional phase space as a manifold $\Gamma \ $%
defined via the following symplectic 2-form $w$\cite{esd1}$,$%
\begin{equation}
w=dq^{\mu }\wedge dp_{\mu };\,\,\mu =1,2,..N,  \label{sympl}
\end{equation}
and by a vector field 
\begin{equation}
X_{f}=\frac{\partial f}{\partial p_{\mu }}\frac{\partial }{\partial q^{\mu }}%
-\frac{\partial f}{\partial q^{\mu }}\frac{\partial }{\partial p_{\mu }},
\label{vector field}
\end{equation}
where $q=(q^{1},...,q^{N})$ = $({\bf q},q^{k+1},...,q^{N})$ is a point of $%
{\cal G}$ (hereafter we use the notation: ${\cal G}_{N,k}$ to indicate the
dimension $N$ of the space and $k$ standing fot the dimension of ${\bf I}%
_{ij}$ in the metric) characterizing the configuration space in generalized
coordinates, and $f$ is a $C^{\infty }$ function in $\Gamma $, such that a
Poisson bracket $\{f,g\}$ is introduced by 
\begin{eqnarray}
w(X_{f},X_{h}) &=&\{f,h\}  \nonumber \\
&=&dq^{\mu }(X_{f})dp_{\mu }(X_{h})-dp_{\mu }(X_{f})dq^{\mu }(X_{h}), 
\nonumber \\
&=&\eta ^{\mu \nu }(\frac{\partial f}{\partial q^{\mu }}\frac{\partial h}{%
\partial p^{\nu }}-\frac{\partial f}{\partial p^{\nu }}\frac{\partial h}{%
\partial q^{\mu }}).  \label{pp3}
\end{eqnarray}
On the other hand we have 
\begin{equation}
\{f,h\}=\text{d}f(X_{h})=\left\langle \text{d}f,X_{h}\right\rangle .
\label{c18}
\end{equation}
and 
\[
\left\langle \text{d}f,X_{\{h,g\}}\right\rangle +\left\langle \text{d}%
h,X_{\{g,f\}}\right\rangle +\left\langle \text{d}g,X_{\{f,h\}}\right\rangle
=0,
\]
such that 
\begin{equation}
\lbrack X_{h},X_{f}]=X_{\{f,h\}},  \label{c17}
\end{equation}
inducing then a representation of Lie groups. Considering a Lie algebra
characterized by the structure constants $C_{ij}^{k}\ $, the canonical
representation can be introduced by following usual methods, that is, 
\begin{equation}
\{l_{i},l_{j}\}=C_{ij}^{k}l_{k},  \label{pp4}
\end{equation}
where $l_{j}$ are functions in $\Gamma $ corresponding to generators of the
Lie symmetries. A canonical representation for such Lie algebra, given by
Eqs.(\ref{comm1})-(\ref{comm3}), but with Lie product defined by (\ref{pp3})
as in (\ref{pp4}), is provided by the generators, 
\begin{equation}
M_{\mu \nu }=q_{\mu }p_{\nu }-p_{\mu }q_{\nu },\,\,\text{and\thinspace
\thinspace \thinspace }P_{\mu }=p_{\mu }.  \label{mom1}
\end{equation}

On the other hand, the vector fields $X_{f}$ and the functions $f$ induce a
unitary representation in the Hilbert space ${\cal H}(\Gamma )$ built up
from the complex functions in $\Gamma $ of type $L^{2}$(Lebesgue integral).
Actually, from an arbitrary function $f$ we can construct unitary operators
from the vector field, that is $\widehat{f}=-iX_{f}$. It is worth to note
that $f$ also induces multiplicative operators of ${\cal H}(\Gamma )$ given
by $\overline{f}={\bf 1}f.$ As a consequence we have 
\begin{eqnarray}
\lbrack \widehat{M}_{\mu \nu },\widehat{M}_{\rho \sigma }] &=&-i\left( \eta
_{\nu \rho }\widehat{M}_{\mu \sigma }-\eta _{\mu \rho }\widehat{M}_{\nu
\sigma }+\eta _{\mu \sigma }\widehat{M}_{\nu \rho }-\eta _{\nu \sigma }%
\widehat{M}_{\mu \rho }\right) ,  \label{c30} \\
\lbrack \widehat{P}_{\mu },\widehat{M}_{\rho \sigma }] &=&-i\left( \eta
_{\mu \rho }\widehat{P}_{\sigma }-\eta _{\mu \sigma }\widehat{P}_{\rho
}\right) ,  \label{c333} \\
\lbrack \widehat{P}_{\mu },\widehat{P}_{\rho }] &=&0,  \label{c334}
\end{eqnarray}
with the auxiliary extra conditions 
\begin{eqnarray}
\lbrack \widehat{M}_{\mu \nu },M_{\rho \sigma }] &=&-i\left( \eta _{\nu \rho
}M_{\mu \sigma }-\eta _{\mu \rho }M_{\nu \sigma }+\eta _{\mu \sigma }M_{\nu
\rho }-\eta _{\nu \sigma }M_{\mu \rho }\right) ,  \label{c35} \\
\lbrack \widehat{P}_{\mu },M_{\rho \sigma }] &=&-i\left( \eta _{\mu \rho
}P_{\sigma }-\eta _{\mu \sigma }P_{\rho }\right) ,  \label{c36} \\
\lbrack \widehat{P}_{\mu },P_{\rho }] &=&0,  \label{c37}
\end{eqnarray}
where $M_{\rho \sigma }$ and $P_{\sigma }$ are given in Eq. (\ref{mom1}) and 
$\widehat{M}_{\mu \nu }$ is 
\begin{equation}
\widehat{M}_{\mu \nu }=ip_{\nu }\frac{\partial }{\partial p^{\mu }}-ip_{\mu }%
\frac{\partial }{\partial p^{\nu }}+iq_{\nu }\frac{\partial }{\partial
q^{\mu }}-iq_{\mu }\frac{\partial }{\partial q^{\nu }};\;\;\widehat{P}_{\mu
}=-i\frac{\partial }{\partial q^{\mu }}.  \label{mom2}
\end{equation}

We remark introducing unitary operators defined by 
\[
\widehat{f}=f-\frac{1}{2}(q^{\mu }\frac{\partial f}{\partial q^{\mu }}%
+p^{\mu }\frac{\partial f}{\partial p^{\mu }})-iX_{f},
\]
which is a generalization of the van Hove bracket \cite{van Hove}, Eqs. (\ref
{c30})-(\ref{c37}) are derived with $\widehat{M}_{\rho \sigma }$ given again
by (\ref{mom2}) and 
\begin{equation}
\widehat{P}_{\mu }=\frac{1}{2}p_{\mu }-i\frac{\partial }{\partial q^{\mu }}.
\label{mom3}
\end{equation}

In the next section we use the two embeddings treated at the end of Section
2 in association with this phase space approach.

\section{ Applications in kinetic theory}

\subsection{Poincar\'e Group in phase space}

For the embedding described by Eq.(\ref{emb1}), we have for $N=5,$ $\Im _{1}:%
{\bf q}\ \mapsto \ q=({\bf q},\frac{q_{4}}{\sqrt{2}},\frac{q_{4}}{\sqrt{2}}%
), $ and $\Im _{1}:{\bf p}\ \mapsto \ p=({\bf p},\frac{p_{4}}{\sqrt{2}},%
\frac{p_{4}}{\sqrt{2}})$, with $(q,p)\,\in \Gamma .$ So doing, taking into
account Eq.(\ref{mink11}), we get 
\[
q^{\mu }q_{\mu }=\eta ^{\mu \nu }q_{\mu }q_{\nu }\stackrel{\Im _{1}}{\mapsto 
}q^{\mu }q_{\mu }=g^{\mu \nu }q_{\mu }q_{\nu } 
\]
being now $\mu ,\nu =0,1,2,3$ in $g^{\mu \nu }.$ We have with this embeding, 
$M_{\mu \nu },$\thinspace \thinspace $\widehat{M}_{\mu \nu }$ and $\widehat{P%
}_{\mu }$ ($\mu ,\nu =0,1,2,3)$ given formally by Eqs. (\ref{mom1}), (\ref
{mom2}) and (\ref{mom3}), and $P_{\mu }=p_{\mu }\stackrel{\Im _{1}}{\mapsto }%
(p_{0,}p_{1},p_{2},p_{3}).$ Eqs.(\ref{c30})-(\ref{c37}) are then replaced by 
\begin{eqnarray}
\lbrack \widehat{M}_{\mu \nu },\widehat{M}_{\rho \sigma }] &=&-i\left(
g_{\nu \rho }\widehat{M}_{\mu \sigma }-g_{\mu \rho }\widehat{M}_{\nu \sigma
}+g_{\mu \sigma }\widehat{M}_{\nu \rho }-g_{\nu \sigma }\widehat{M}_{\mu
\rho }\right) ,  \label{cc30} \\
\lbrack \widehat{P}_{\mu },\widehat{M}_{\rho \sigma }] &=&-i\left( g_{\mu
\rho }\widehat{P}_{\sigma }-g_{\mu \sigma }\widehat{P}_{\rho }\right) ,
\label{cc333} \\
\lbrack \widehat{P}_{\mu },\widehat{P}_{\rho }] &=&0,  \label{cc334}
\end{eqnarray}
and 
\begin{eqnarray}
\lbrack \widehat{M}_{\mu \nu },M_{\rho \sigma }] &=&-i\left( g_{\nu \rho
}M_{\mu \sigma }-g_{\mu \rho }M_{\nu \sigma }+g_{\mu \sigma }M_{\nu \rho
}-g_{\nu \sigma }M_{\mu \rho }\right) ,  \label{cc35} \\
\lbrack \widehat{P}_{\mu },M_{\rho \sigma }] &=&-i\left( g_{\mu \rho
}P_{\sigma }-g_{\mu \sigma }P_{\rho }\right) ,  \label{cc36} \\
\lbrack \widehat{P}_{\mu },P_{\rho }] &=&0.  \label{cc37}
\end{eqnarray}
This algebra is a representation of the Poincar\'{e}-Lie algebra in the
relativistic phase space derived in the context of the classical
theormofield dynamics formalism\cite{marco1}. Some Casimir invariants are
given by 
\begin{eqnarray}
I_{1} &=&P^{\mu }P_{\mu },  \label{i1} \\
I_{2} &=&\widehat{P}^{\mu }\widehat{P}_{\mu },  \label{i2} \\
I_{3} &=&P^{\mu }\widehat{P}_{\mu },  \label{i3}
\end{eqnarray}
resulting immediately, from $I_{3},$ in the relativistic kinetic equation 
\begin{equation}
p^{\mu }\partial _{\mu }\phi (q,p)=0  \label{rel eq}
\end{equation}
without the collision term, where $\phi (q,p)$ is a classical amplitude of
probability in phase space \cite{marco1}. The density of probability is
given by $f(q,p)=|\phi (q,p)|^{2}$, still satisfying the same Eq.(\ref{rel
eq}). In Ref.\cite{marco1}, using the notion of propagator, a
phenomenological collision term was derived for Eq.(\ref{rel eq}). Here we
would like to emphasize that Eq.(\ref{rel eq}) can be derived from a
variational principle so that interaction terms can be introduced by
imposing symmetry as a physical criterious, as we proceed in usual field
theory. For instance, from the Lagrangian density 
\begin{equation}
L=\frac{1}{2}\phi ^{\ast }(q,p)p^{\mu }\partial _{\mu }\phi (q,p)+\frac{1}{2}%
\phi (q,p)p^{\mu }\partial _{\mu }\phi ^{\ast }(q,p)-\lambda V(\phi ,\phi
^{\ast }),  \label{lagran1}
\end{equation}
we obtain the Euler-Lagrange equation 
\[
p^{\mu }\partial _{\mu }\phi (q,p)=\lambda \frac{\partial V(\phi ,\phi
^{\ast })}{\partial \phi ^{\ast }}. 
\]
A model for $V(\phi )$ preserving the phase invariance $\phi \rightarrow
-\phi $ is given by $V(\phi )=\frac{1}{4}\phi ^{4},\ $resulting in 
\begin{equation}
p^{\mu }\partial _{\mu }\phi (q,p)=\lambda \phi ^{3}.  \label{aur1}
\end{equation}

To avaliate the interest of this method, let us consider Eq.(\ref{aur1}) in $%
(1+1)$ dimensions, that is 
\begin{equation}
(\partial _{t}+\frac{p^{1}}{p^{o}}\partial _{x})\phi (x,p)=\frac{\lambda }{%
p^{o}}\phi ^{3}.  \label{aur2}
\end{equation}
Assuming a wave-travelling solution, 
\[
\phi (x,p)\equiv \psi (vt+x).
\]
and defining $y=vt+x,$ Eq.(\ref{aur1}) reads 
\[
\partial _{y}\psi =\frac{\lambda }{(vp^{o}+p^{1})}\psi ^{3}.
\]
A solution of this equations is given by 
\[
\phi (x,p)=\psi (x,p)=\sqrt{\frac{vp^{o}+p^{1}}{\lambda (vt+x)}}.
\]
Observe that the probability\ density function in the phase space is
interpreted here by $f(q,p)=|\phi (q,p)|^{2}.$ Therefore we find 
\[
f(x,p)=\frac{vp^{o}+p^{1}}{(vt+x)\lambda }.
\]
It is worthy noticing the fact that this simple distribution function can
not be derived so trivially, as it was here, by using the usual formalism
based on the Boltzmann-like equations for $f(x,p),$ since the collision term
is often given by cumbersome arguments. In the sequence, as another
application, we use the embedding given in Eq.(\ref{emb2}) and the Weyl
operators.

\subsection{Weyl representation for the Electromagnetic waves in dispersive
medium}

McDonald and Kaufman \cite{mck} have derived a wave kinetic equation
governing the behavior of electromagnetic waves in dispersive medium in the
following way. The electric field ${\bf E}({\bf q},t)$ propagating in an
inhomogeneous, time-varying medium can be described by the integral equation 
\begin{equation}
\int d^{3}q^{\prime }dt^{\prime }{\cal D}({\bf q},t;{\bf q}^{\prime
},t^{\prime }){\bf E}({\bf q},t)={\bf j}({\bf x},t).  \label{mc33}
\end{equation}
Eq. (\ref{mc33})\ can be seen as an operator equation given by ${\bf D\cdot
E=j.}$ Using its adjoint counterpart, we can write 
\begin{equation}
{\bf D\cdot }({\bf EE}^{{\bf \dagger }}){\bf =}({\bf jj}^{{\bf \dagger }})(%
{\bf D}^{{\bf \dagger }})^{-1},  \label{disp1}
\end{equation}
which is an equation for the spectral tensor of the electric field $({\bf EE}%
^{{\bf \dagger }})$ written in terms of the dispersion operator ${\bf D}$
and the spectral current-source tensor $({\bf jj}^{{\bf \dagger }}).$ A
phase space representation for Eq. (\ref{disp1}) is then derived by using
the Weyl representation for the product of operators, resulting in \cite{mck}
\begin{equation}
D\Delta ({\bf q},t;{\bf k},w)(EE^{{\bf \dagger }})=(jj^{{\bf \dagger }%
})\Delta ({\bf q},t;{\bf k},w)(D^{{\bf \dagger }})^{-1},  \label{disp22}
\end{equation}
where 
\begin{equation}
\Delta ({\bf q},t;{\bf k},w)=\exp \frac{i}{2}(\frac{\stackrel{\leftarrow }{%
\partial }}{\partial q^{i}}\frac{\stackrel{\rightarrow }{\partial }}{%
\partial k_{i}}-\frac{\stackrel{\leftarrow }{\partial }}{\partial k_{i}}%
\frac{\stackrel{\rightarrow }{\partial }}{\partial q^{i}}+\frac{\stackrel{%
\leftarrow }{\partial }}{\partial w}\frac{\stackrel{\rightarrow }{\partial }%
}{\partial t}-\frac{\stackrel{\leftarrow }{\partial }}{\partial t}\frac{%
\stackrel{\rightarrow }{\partial }}{\partial w}).  \label{disp23}
\end{equation}

The main point here is that we can find a geometric interpretation for the
McDonald and Kaufman theory\cite{mck}. Let us consider the embedding given
in Eq.(\ref{emb2}) in 5-dimensions, such that $\Im _{2}:{\bf q}\ \mapsto \
q=({\bf q},t,0).$ The conjugate variable of $q,$ say $k,$ is of the form $%
{\bf k}\stackrel{\Im _{2}}{\mapsto }\ $ $k=({\bf k},0,w).$ Then, from Eq.(%
\ref{sympl}) we have 
\begin{eqnarray}
w &=&\text{d}q^{\mu }\wedge \text{d}k_{\mu }\   \nonumber \\
\ &=&\text{d}q^{i}\wedge \text{d}k_{i}\ -dt\wedge dw,  \label{c345}
\end{eqnarray}
and

\[
X_{f}=\frac{\partial f}{\partial k^{i}}\frac{\partial }{\partial q_{i}}-%
\frac{\partial f}{\partial q_{i}}\frac{\partial }{\partial k^{i}}+\frac{%
\partial f}{\partial w}\frac{\partial }{\partial t}-\frac{\partial f}{%
\partial t}\frac{\partial }{\partial w}. 
\]
such that 
\begin{eqnarray}
w(X_{f},X_{h}) &=&\{f,h\}  \nonumber \\
&=&dq^{\mu }(X_{f})dp_{\mu }(X_{h})-dp_{\mu }(X_{f})dq^{\mu }(X_{h}) 
\nonumber \\
&=&\eta ^{\mu \nu }(\frac{\partial f}{\partial q^{\mu }}\frac{\partial h}{%
\partial p^{\nu }}-\frac{\partial f}{\partial p^{\nu }}\frac{\partial h}{%
\partial q^{\mu }})  \nonumber \\
&=&\frac{\partial f}{\partial q^{i}}\frac{\partial h}{\partial k_{i}}-\frac{%
\partial f}{\partial k_{i}}\frac{\partial h}{\partial q^{i}}+\frac{\partial f%
}{\partial w}\frac{\partial h}{\partial t}-\frac{\partial f}{\partial t}%
\frac{\partial h}{\partial w}.  \label{disp25}
\end{eqnarray}

With this Poisson bracket, we construct a Weyl product of functions in phase
space $f(q,p)$ and $\,g(q,p)$ (interpreted as a phase space representation
of a product of two corresponding operators $F\cdot G)$ by 
\begin{equation}
f\ast g=f(q,p)\,\Delta ({\bf q},t;{\bf k},w)\,g(q,p).  \label{we1}
\end{equation}
where $\Delta ({\bf q},t;{\bf k},w)$ is given, from Eq.(\ref{disp25}), in
Eq.(\ref{disp23}).

\section{The connection between the \ CJ manifold and the ARS-group}

Let us consider the ${\cal G}$ manifold in six dimensions, with $N=6$ and $%
k=2$ (${\cal G}_{6,2}$). The metric $\eta $ is then 
\begin{equation}
\left( \eta _{\mu \nu }\right) =\left( 
\begin{array}{ccc}
{\bf I}_{2} & 0 & 0 \\ 
0 & {\Bbb B}_{1} & 0 \\ 
0 & 0 & {\Bbb B}_{2}
\end{array}
\right) ,  \label{met-6}
\end{equation}
where $diag\left( {\bf I}_{2}\right) =\left( 1,1\right) ${\bf \ }and{\bf \ }$%
{\Bbb B}_{i}$ is given by the Eq. (\ref{block}). Let us denote a vector in $%
{\cal G}_{6,2}$ by $\xi =\left( \xi ^{1},...,\xi ^{6}\right) $. Through a
mapping, $\eta $ can made partially diagonal, that is introduce the
transformation $U:{\cal G}_{6,2}\rightarrow {\cal G}_{6,2},$ 
\begin{eqnarray}
\xi ^{i} &\mapsto &\xi ^{i},\;  \nonumber \\
U &:&\xi ^{3}\mapsto \frac{1}{\sqrt{2}}\left( \xi ^{4}-\xi ^{3}\right) , 
\nonumber \\
U &:&\xi ^{4}\mapsto \frac{1}{\sqrt{2}}\left( \xi ^{4}+\xi ^{3}\right) ,
\label{new var} \\
U &:&\xi ^{5}\mapsto \xi ^{5},  \nonumber \\
U &:&\xi ^{6}\mapsto \xi ^{6}.  \nonumber
\end{eqnarray}
These relations can be rewritten as 
\begin{equation}
x^{\mu }=U_{v}^{\mu }\xi ^{\nu }\;  \label{new var 2}
\end{equation}
where $\mu ,v=1,\ldots ,6,$ the $x^{\mu }$ are the new variables, and $%
U_{v}^{\mu }$ are the transformation coefficients given by 
\begin{equation}
\left( U_{v}^{\mu }\right) =\left( 
\begin{array}{cccc}
{\bf I}_{2} & 0 & 0 & 0 \\ 
0 & -\frac{1}{\sqrt{2}} & \frac{1}{\sqrt{2}} & 0 \\ 
0 & \frac{1}{\sqrt{2}} & \frac{1}{\sqrt{2}} & 0 \\ 
0 & 0 & 0 & {\bf I}_{2}
\end{array}
\right) ,  \label{transf new var}
\end{equation}
with $U^{-1}=U$.

Considering vectors $\xi ,\zeta \in {\cal G}_{6,2}$, and $x=U\xi $ and $%
y=U\zeta ,$ the inner product between them is 
\begin{eqnarray}
\left( \xi \mid \zeta \right)  &=&\eta _{\mu \nu }\xi ^{\mu }\zeta ^{v} 
\nonumber \\
&=&\left( U_{\alpha }^{\mu }\eta _{\mu \nu }U_{\beta }^{v}\right) x^{\alpha
}y^{\beta }  \nonumber \\
&=&\omega _{\alpha \beta }x^{\alpha }y^{\beta }.  \label{inner pro}
\end{eqnarray}
where $\omega _{\alpha \beta }=U_{\alpha }^{\mu }\eta _{\mu \nu }U_{\beta
}^{v},$ or explicitly, 
\begin{equation}
\left( \omega _{\alpha \beta }\right) =\left( 
\begin{array}{ccc}
g_{\mu \nu } & 0 & 0 \\ 
0 & 0 & -1 \\ 
0 & -1 & 0
\end{array}
\right) ,  \label{new metric exp}
\end{equation}
with $diag\left( g_{\mu \nu }\right) =\left( 1,1,1,-1\right) $ being the $%
(3+1)$ Minkowski metric; Then, we can write Eq.(\ref{inner pro}) as 
\begin{equation}
\left( \xi \mid \zeta \right) =\left( x\mid y\right) ={\bf x}\cdot {\bf y}%
-x^{4}y^{4}-x^{5}y^{6}-x^{6}y^{5}.  \label{new inner pro}
\end{equation}

Analyzing linear transformations of type $\overline{\xi }^{\mu }=G_{\ \nu
}^{\mu }\xi ^{\nu }+\chi ^{\mu }$, such that $\left| G\right| =1,$ with the
use of Eq.(\ref{new var 2}) we obtain 
\begin{equation}
\overline{x}^{\mu }=\widetilde{G}_{\ \nu }^{\mu }x^{\nu }+a^{\mu },
\label{new iso transf}
\end{equation}
where $\widetilde{G}_{\ \nu }^{\mu }=U_{\alpha }^{\mu }G_{\ \beta }^{\alpha
}U_{v}^{\beta }.$ Then the generators of these transformations described by
Eq.(\ref{new iso transf}) are 
\begin{equation}
\widetilde{M}_{\mu \nu }=U_{\mu }^{\alpha }M_{\alpha \beta }U_{v}^{\beta },\;%
\widetilde{A}_{\mu }=U_{\mu }^{\alpha }A_{\alpha },  \label{new generator}
\end{equation}
with $\widetilde{M}$ standing for the rotations and $\widetilde{A}$ for
translations. Hence, we get the following Lie algebra 
\begin{eqnarray}
\left[ \widetilde{M}_{\mu \nu },\widetilde{M}_{\rho \sigma }\right]
&=&-i\left( \omega _{\nu \rho }\widetilde{M}_{\mu \sigma }-\omega _{\mu \rho
}\widetilde{M}_{\nu \sigma }+\omega _{\mu \sigma }\widetilde{M}_{\nu \rho
}-\omega _{\nu \sigma }\widetilde{M}_{\mu \rho }\right) ,
\label{algebraG6-1} \\
\left[ \widetilde{A}_{\mu },\widetilde{M}_{\rho \sigma }\right] &=&-i\left(
\omega _{\mu \rho }\widetilde{A}_{\sigma }-\omega _{\mu \sigma }\widetilde{A}%
_{\rho }\right) ,  \label{algebraG6-2} \\
\left[ \widetilde{A}_{\mu },\widetilde{A}_{\nu }\right] &=&0.
\label{algebraG6-3}
\end{eqnarray}

Let us introduce the following notation for the generators in Eq.(\ref{new
generator}), 
\begin{eqnarray}
\widetilde{M}_{\mu \nu } &=&-\widetilde{M}_{\nu \mu }:=J_{\mu \nu },\;\;%
\widetilde{M}_{5\mu }=-\widetilde{M}_{\mu 5}:=Q_{\mu },  \label{ge2} \\
\widetilde{M}_{6\mu } &=&-\widetilde{M}_{\mu 6}:=N_{\mu },\;\;\widetilde{M}%
_{56}=-\widetilde{M}_{65}:=D,  \label{ge4} \\
\widetilde{A}_{\mu } &:&=P_{\mu },\;\;\widetilde{A}_{5}:=P_{5},\;\;%
\widetilde{A}_{6}:=S,  \label{ge7}
\end{eqnarray}
where now the indices $\mu ,\nu $ run as $\mu ,\nu =1,2,3,4.$ Using this
indices, Eqs.(\ref{algebraG6-1})-(\ref{algebraG6-3}) can then be rewritten
in the $(3+1)$ Minkowski space, that is 
\begin{eqnarray}
\left[ J_{\mu \nu },J_{\rho \sigma }\right]  &=&-i\left( g_{\nu \rho }J_{\mu
\sigma }-g_{\mu \rho }J_{\nu \sigma }+g_{\mu \sigma }J_{\nu \rho }-g_{\nu
\sigma }J_{\mu \rho }\right) ,  \label{alg new group-1} \\
\left[ P_{\mu },J_{\rho \sigma }\right]  &=&-i\left( g_{\mu \rho }P_{\sigma
}-g_{\mu \sigma }P_{\rho }\right) ,\;\;\;\;  \label{alg new group-2} \\
\lbrack P_{\mu },P_{\nu }] &=&[Q_{\mu },Q_{\nu }]=[J_{\mu \nu },S]=[P_{\mu
},S]=0,  \label{alg new group-3} \\
\lbrack Q_{\mu },J_{\rho \sigma }] &=&-i\left( g_{\mu \rho }Q_{\sigma
}-g_{\mu \sigma }Q_{\rho }\right) ,  \label{alg new group-4} \\
\lbrack P_{\mu },Q_{\nu }] &=&ig_{\mu \nu }P_{5},\;\;\;[S,Q_{\mu }]=iP_{\mu
},  \label{alg new group-5} \\
\lbrack P_{5},J_{\rho \sigma }] &=&[P_{5},Q_{\mu }]=[P_{5},P_{\mu
}]=[P_{5},S]=0,\;  \label{alg new group-6}
\end{eqnarray}
We see by means of Eqs.(\ref{alg new group-1})-(\ref{alg new group-6}) that
the generators $J,P,Q\ $ and $S$ form a closed algebra, which is nothing but
the Lie algebra of the called extended relativistic Galilei group, or $ARS-$%
group \cite{Santilli1,Santilli3,Santilli2}. The operators $J_{\mu \nu }$ and 
$P_{\mu }$ stand for the generators of the Poincar\'{e} group, $Q_{\mu }$
are the generators of the \ ``relativistic Galilean boost'', $S$ is the
generator of translations in the $x^{6}$ coordinate, interpreted as the
proper time, and $P_{5}$ is associated with $l^{-1}{\bf 1}$, where $l$ is a
constant with dimension of length. Following this approach, the particular
transformations generated by $J,P,Q\ $ and $S$ are, from Eq.(\ref{new iso
transf}), of the form 
\begin{eqnarray}
\ {\bar{x}}^{\mu } &=&\Lambda _{\ \nu }^{\mu }x^{\nu }+b^{\mu }x^{6}+a^{\mu
},  \label{sant transf1} \\
{\bar{x}}^{5} &=&x^{5}+\left( \Lambda _{\ \nu }^{\mu }x^{\nu }\right) b_{\mu
}+\frac{1}{2}b^{\mu }b_{\mu }x^{6}+a^{5},  \label{sant transf2} \\
{\bar{x}}^{6} &=&x^{6}+a^{6}.  \label{sant transf3}
\end{eqnarray}

These Eqs.(\ref{sant transf1})--(\ref{sant transf3}) are identified as the
transformations of the ARS-group, with $\Lambda _{\,\,\,\,\,\,\nu }^{\mu }$\
representing the $SO(3,1),$ $b^{\mu }$ standing for the relativistic
Galilean boost, and $a^{\mu }$ for the translations. The name ``relativistic
Galilei group'' is justified by the likeness between it and \ the Galilei
group, although it describes relativistic physics. Eq.(\ref{sant transf2})
arises as a compatibility condition, because it gives the isometry of the
transformations in ${\cal G}_{6,2}.$ The generator $P_{5},$ which is here an
invariant in the subalgebra generated by $J,P,Q\ $ and $S,$ is considered in
the $ARS-$approach as a central extension of the group given in Eqs.\bigskip
(\ref{sant transf1}) and Eq.(\ref{sant transf3}).

In order to see the compatibility of this $ARS-$formalism with the usual
one, let us derive the scalar representation but from the CJ manifold
context, that is using unitary representation for the linear transformations
in ${\cal G}_{6,2}$. From Eqs.(\ref{alg new group-1})-(\ref{alg new group-6}%
) two of the Casimir invariants are given by 
\begin{eqnarray}
I_{1} &=&P_{\mu }P^{\mu },  \label{Cas 1} \\
I_{2} &=&P_{5}.  \label{Cas 2}
\end{eqnarray}
These invariants $I_{1}$\ and $I_{2}$ can be used to get physical
representations of the $ARS-$group. Considering the simplest case, i.e. a
faithful unitary representation in the Hilbert space ${\cal H}\left( {\cal G}%
_{6,2}\right) $ of scalar functions with $I_{1}=-\partial _{\mu }\partial
^{\mu }$ and $I_{2}=-i\partial _{5}$, we have from the Schur%
\'{}%
s Lemma, 
\begin{eqnarray}
-\partial _{\mu }\partial ^{\mu }\Psi  &=&k\Psi ,  \label{first eq} \\
-i\partial _{5}\Psi  &=&l^{-1}\Psi ,  \label{sec eq}
\end{eqnarray}
where $\Psi =\Psi \left( x^{1},...,x^{6}\right) $ is a scalar function, and $%
k$ and $l$ are arbitrary constants. So, substituting Eq.(\ref{sec eq}) in
Eq.(\ref{first eq}) and setting $k=0,$ we get 
\begin{equation}
\left( \nabla ^{2}-\partial _{4}^{2}-2il^{-1}\partial _{6}\right) \Psi =0.
\label{thrird eq}
\end{equation}
Fixing on functions of type $\Psi \left( x\right) =\Phi \left(
x^{1},x^{2},x^{3},x^{4}\right) \chi \left( x^{5},x^{6}\right) $ and $\frac{1%
}{2}im^{2}l$ as the separation constant, we obtain 
\begin{equation}
\left( \nabla ^{2}-\partial _{4}^{2}+m^{2}\right) \Phi =0.  \label{K-G eq}
\end{equation}
Hence, identifying $\partial _{4}\equiv \partial _{t},$ $m$ being the mass
and $\Phi $ the wave function of the particle, Eq.(\ref{K-G eq}) is just the
Klein-Gordon equation.

\section{Concluding Remarks}

In this work we have explored the Cangemi-Jackiw space (${\cal G}$), a
generalized light-cone manifold. Tensor fields and symplectic structures
have been introduced for arbitrary dimensions. In particular, canonical
representations of Lie symmetry are defined. Through the notion of
embeddings of the Euclidean space into ${\cal G}$ and exploring methods of
field theory, we have derived a representation for the Poincar\'{e} group in
phase space which is closely associated with the relativistic kinetic
theory. Using then arguments based on symmetry and covariance, a Lagrangian
with an interaction term is written in phase-space. The Euler-Lagrange
equations give rise to a model-dependent kinetic equation, which is a
counterpart in phase space of the usual $\lambda \phi ^{4}$ theory. A
solution for this equation provides a travelling-wave distribution function.
The simplicity in the reasoning leading to such a result lies on two
ingredients (which are not usual in phase-space theories): the notion of
amplitude and a wcovariant Lagrangian. In this context, different
possibilities for the potential $V(\phi )$ in Eq.(\ref{lagran1}), as for
instance $V(\phi )=k\cos (\beta \phi ),$ remain to be explored in future
works.

The formalism has also been used to provide geometrical basis for the
McDonald and Kaufman approach \cite{mck}, which describes the behavior of
the electromagnetic field in dispersive medium. This interpretative result
opens a possibility for using their method, under this new geometrical
understanding, to consider other complex situations, such as the derivation
of kinetic equations for non-abelian fields. Finally, we have employed $%
{\cal G}$ in 6-dimensions to derive the Aghassi-Roman-Santilli\ group \cite
{Santilli1,Santilli3,Santilli2}, which is a kind of Galilei group defined in
the $(3+1)$-Minkowski space. With these applications the remarkable
quiescence of the CJ manifold is made more evident.

\begin{description}
\item  {\bf Acknowledgments }We would like to thank J.D.M. Vianna, F. C.
Khanna, M. de Montigny and E. S. Santos for interesting discussions, and
CNPq and CAPES (two Brazilian Government Agencies) for financial support.
\end{description}


\begin{references}
\bibitem{cj1}  D. Cangemi and R. Jackiw, Phys, Rev. Lett. {\bf 69} (1992)
233.

\bibitem{cj2}  D. Cangemi and R. Jackiw, Phys. Lett. B {\bf 299} (1993) 24.

\bibitem{cj3}  D. Cangemi and R. Jackiw, Ann. Phys. (N.Y.) {\bf 225} (1993)
229.

\bibitem{Bazeia}  D. Bazeia and R. Jackiw, Ann. Phys. (N.Y.) {\bf 270}
(1998) 246.

\bibitem{Susskind}  L. Susskind, Phys. Rev. {\bf 165} (1968)\ 1535.

\bibitem{wzw}  C. R. Nappi and E. Witten, Phys. Rev. Lett. {\bf 71} (1993)
3751; W. Witten, Comm. Math. Phys. {\bf 92} (1984) 455.

\bibitem{olive}  D.I. Olive, E. Rabinovici, and A Schwimmer, Phys. Lett.B 
{\bf 321} (1994) 361;

\bibitem{sf}  K. Sfetsos, Phys. Lett.B {\bf 324} (1994) 335.

\bibitem{luk}  J. Lukierski, P.C. Stichel, and W.J. Zakrzewski, Ann. Phys. 
{\bf 260} (1997) 224.

\bibitem{esd1}  M. de Montigny, F. C. Khanna, A. E. Santana, E. S. Santos,
J. D. M. Vianna, Ann. Phys (N.Y.) {\bf 277} (1999) 144.

\bibitem{kun1}  C. Duval, G. Burdet, H.P. K\"{u}nzle, and M. Perrin, Phys.
Rev. D {\bf 31} (1985) 1841.

\bibitem{kun2}  H. P. K\"unzle, Canad. J. Phys. {\bf 64} (1986) 185.

\bibitem{kun3}  H. P. K\"unzle and C. Duval, Clas. Quant. Grav. {\bf 3}
(1986) 957.

\bibitem{kun4}  H. P. K\"unzle and C. Duval, {\it Semantical aspects of
spacetime theories, }U. Majer and H.-J. Schmidt, eds.
(BI-Wissenschaftsverlag, Mannhein, 1994) p. 113.

\bibitem{tak1}  Y. Takahashi, Fortschr. Phys. {\bf 36} (1988) 63.

\bibitem{tak2}  Y. Takahashi, Fortschr. Phys. {\bf 36} (1988) 83.

\bibitem{tak3}  M. Omote, S. Kamefuchi, Y. Takahashi and Ohnuki, Fortschr.
Phys. {\bf 37} (1989) 933.

\bibitem{tak4}  A.E. Santana, F.C. Khanna, and Y. Takahashi, Prog. Theor.
Phys. {\bf 99} (1998) 327.

\bibitem{van Hove}  L. Van Hove, Proc. R. Acad. Sci. {\bf 26} (1951) 1.

\bibitem{marco1}  M. C. B. Andrade, A. E. Santana, J. D. M. Vianna, J. Phys.
A: Math. Gen. {\bf 33 }(2000) 4015.

\bibitem{Wigner}  E. P. Wigner, Phys. Rev. {\bf 40} (1932) 749.

\bibitem{mck}  S. W. McDonald and A. N. Kaufman, Phys. Rew. A {\bf 32}
(1985) 1708.

\bibitem{Santilli1}  J. J. Aghassi, P. Roman and R. M. Santilli, Phys. Rev.
D {\bf 1 }(1970) 2753.

\bibitem{Santilli3}  J. J. Aghassi, P. Roman and R. M. Santilli, J. Math.
Phys. {\bf 11} (1970) 2297.

\bibitem{Santilli2}  J. J. Aghassi, P. Roman and R. M. Santilli, N. Cimento 
{\bf 5 }(1971) 551.
\end{references}
\end{document}